\documentclass[11pt,a4paper]{article}
\pdfoutput=1
\usepackage{jheppub}

\usepackage{color}
\usepackage{amsmath}
\usepackage{pifont}

\usepackage{bbm}
\usepackage{verbatim}   
\usepackage{subfigure}  
\usepackage{acronym}

\usepackage{amsfonts}
\usepackage{amssymb}
\usepackage{mathrsfs}
\usepackage{graphicx}
\usepackage{multirow}
 \usepackage{slashed}

\newcommand{\MeV}{{\, {\rm MeV}}}
\newcommand{\GeV}{{\, {\rm GeV}}}
\newcommand{\TeV}{{\, {\rm TeV}}}

\def\beq{\begin{equation}}
\def\eeq{\end{equation}}
\def\bea{\begin{eqnarray}}
\def\eea{\end{eqnarray}}
\def\bitem{\begin{itemize}}
\def\eitem{\end{itemize}}
\newcommand{\bec}{\begin{center}}
\newcommand{\eec}{\end{center}}
\newcommand{\ba}{\begin{array}}
\newcommand{\ea}{\end{array}}

\def\bar#1{\overline{#1}}

\def\inv{^{\raise.15ex\hbox{${\scriptscriptstyle -}$}\kern-.05em 1}}
\def\lbar{{\lower.35ex\hbox{$\mathchar'26$}\mkern-10mu\lambda}} 

\let\<=\langle
\let\>=\rangle

\let\+=\uparrow

\def\ra{\rightarrow}

\def\l{\left}
\def\r{\right}
\def \dblarrow#1{\stackrel{\leftrightarrow}{#1}}
\def\dblone{\hbox{$1\hskip -1.2pt\vrule depth 0pt height 1.6ex width 0.7pt 
\vrule depth 0pt height 0.3pt width 0.12em$}}

\begin{document}

\title{Diphotons from Diaxions
} 
\author[a]{Luis Aparicio} \author[a]{, Aleksandr Azatov} \author[a]{, Edward Hardy}  \author[a,b]{, and Andrea Romanino}
\emailAdd{laparici@ictp.it}
\emailAdd{aazatov@ictp.it}
\emailAdd{ehardy@ictp.it}
\affiliation[a]{Abdus Salam International Centre for Theoretical Physics,
Strada Costiera 11, 34151, Trieste, Italy}
\affiliation[b]
{SISSA/ISAS and INFN, 34136 Trieste, Italy}

\abstract{

We study models in which the 750 GeV diphoton excess is due to a
scalar decaying to two pseudo-Nambu-Goldstone bosons that subsequently decay into two pairs of
highly boosted photons, misidentified as individual photons. Performing a model
independent analysis we find that, with axion mass around $200$ MeV, this class of theories can
naturally explain the observed signal with a large width, without violating monojet
constraints. At the same time the requirement of a prompt axion decay can be
satisfied only with a relatively large axion-photon coupling, leading to many new charged particles at the TeV scale. However, the required multiplicities of such fields is still
moderately reduced relative to models with direct diphoton production.

}

\maketitle


\section{Introduction} \label{sec:intro}
Recently both the ATLAS and CMS collaborations observed an excess in diphoton events with an invariant mass in the region of $750\,\GeV$ \cite{atlas,CMS:2015dxe}, and the particularly interesting feature that it appears to have a large width $\Gamma\sim 40 \,\GeV$.

\begin{figure}
\begin{center}
\includegraphics[scale=0.6]{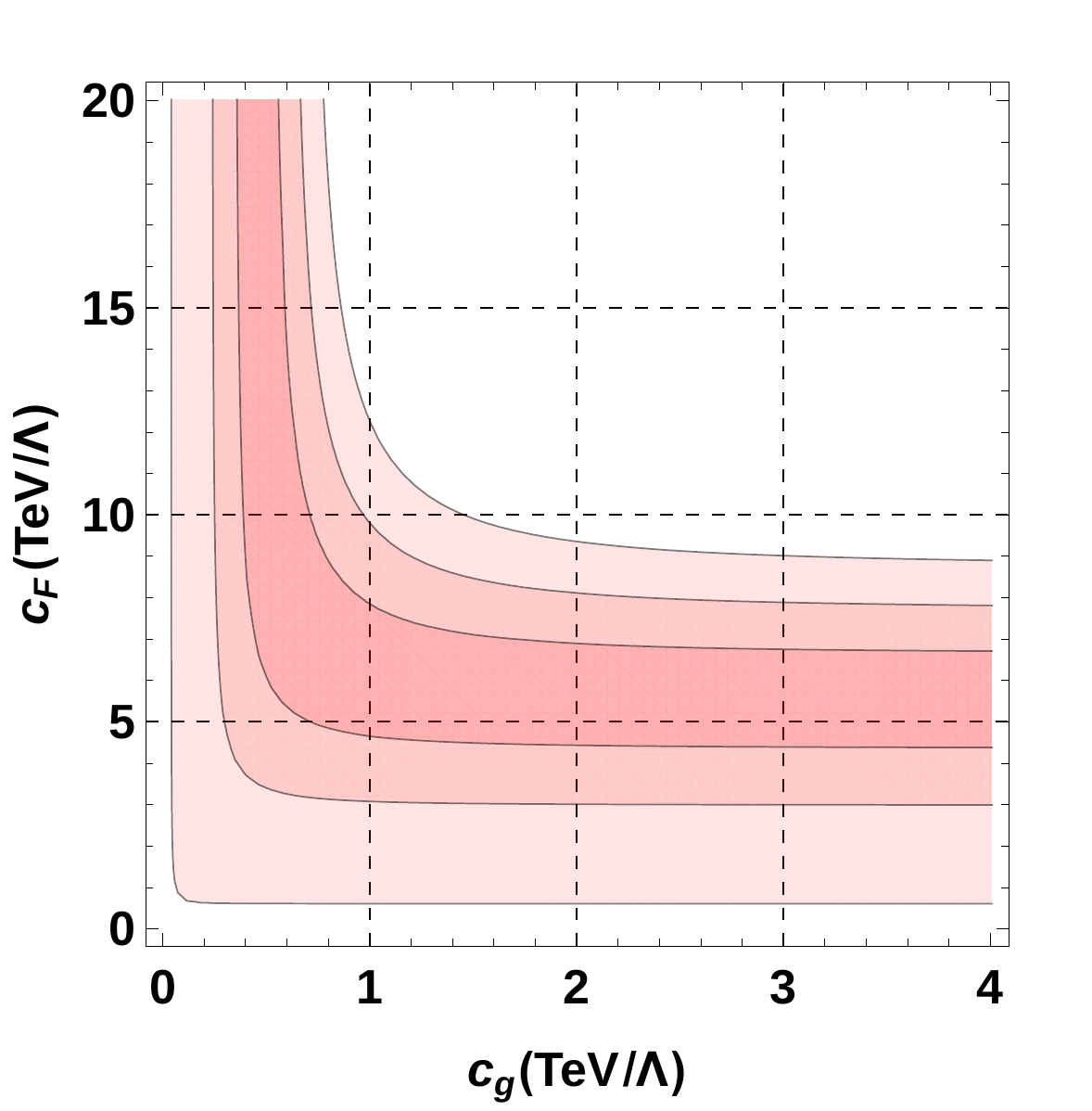}
\includegraphics[scale=0.66]{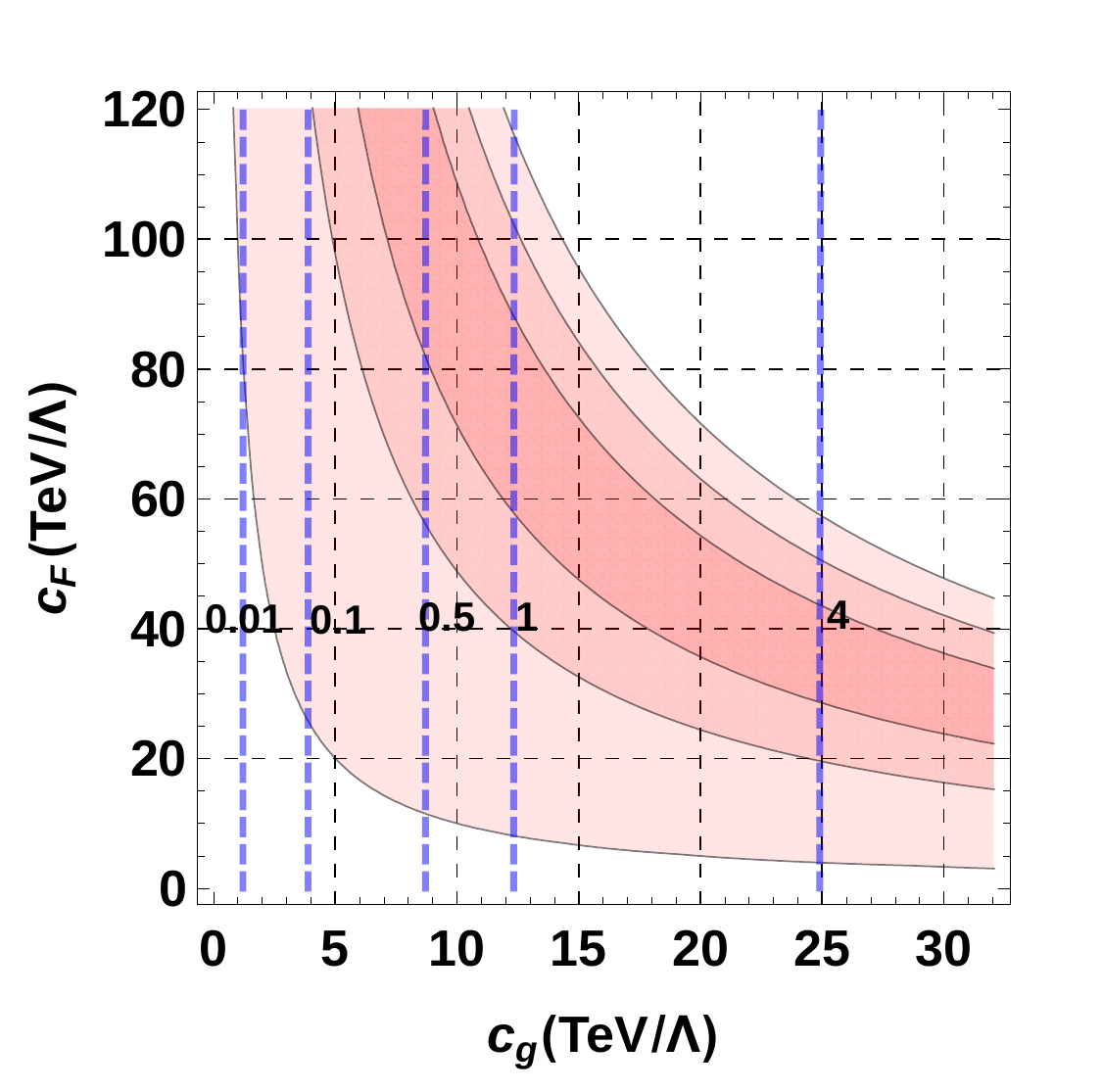}
\end{center}
\caption{\label{fig:noax}{{\bf Left:} $68,95,99 \%$ contours on the fit of the parameterisation Eq.~\eqref{eq:noax} to the ATLAS excess without imposing the width constraint. {\bf Right:}
 $68,95,99 \%$ contours on the same fit when the total  
 width is fixed to be $40$ GeV by the addition of invisible decay channels. Blue lines indicate the corresponding $\sigma(pp \ra$ invisible) cross sections at 8 Tev in pb. Since monojet searches constrain the invisible cross section $\lesssim 0.5\,{\rm pb}$, it is difficult to obtain a large width in this model.
 }}
\end{figure}

The excess has attracted an enormous amount of attention from the theoretical community \cite{blob}. A signal at $13\,\TeV$ can be compatible with the previous $8 \,\TeV$ run if the production of a new scalar resonance is assumed to occur through gluon fusion. This mechanism has a $\sim 4.5$ difference in parton luminosities for 8 and 13  TeV collisions, allowing both measurements to be reconciled with a tension of $\lesssim 2\sigma$. The simplest way to parameterise such a resonance is 
\bea
\label{eq:noax}
{\cal L}=\frac{c_g\alpha_s}{12\pi}\frac{s}{\Lambda}G_{\mu\nu}G^{\mu\nu}+\frac{2\alpha}{9\pi}c_F\frac{s}{\Lambda}F_{\mu\nu}F^{\mu\nu} ~,
\eea
where the couplings are normalised so that $c_g=c_F=1$ corresponds to the  one loop effects of a heavy analogue of the top quark obtaining mass from symmetry breaking at a scale $\Lambda$.  Fitting the signal we obtain
the well known constraints, on $c_F$ and $c_g$ in our case, shown in Fig.~\ref{fig:noax}.
New colour charged states are expected to appear close to $\Lambda$, and LHC constraints require that these have a mass $\gtrsim \TeV$, so the number of observed events can be fit with $O(\hbox{few})$ multiplicities  of new coloured and charged fields. 

Furthermore, the current measurement hints that the signal might be better described by a resonance with width $\sim 40$ GeV. Even though the significance of the currently observed width is very small, for the purpose of our present work we will assume that this feature is required. In the effective model of Eq.~\eqref{eq:noax} the partial decay width into photons and the gluons is given by
 \bea
 \Gamma(s\ra \gamma\gamma, gg)\sim1.6\times 10^{-3} \l(\frac{c_g\hbox{TeV}}{\Lambda}\r)^2+10^{-5}\l(\frac{c_{F} \hbox{TeV}}{\Lambda}\r)^2 \,\GeV ~,
 \eea
 which means $c_g \l({\rm TeV}/\Lambda\r)\gtrsim 160$ or $c_f\l({\rm TeV}/\Lambda\r)\gtrsim 2000$ is needed to have $ \Gamma(s\ra \gamma\gamma, gg)\sim 40 \,\GeV$. However such large values of the couplings are incompatible with current observations: due to either the additional photon production \cite{Fichet:2015vvy,Franceschini:2015kwy,Csaki:2015vek,Csaki:2016raa} or constraints from dijet searches \cite{Franceschini:2015kwy}. Even allowing invisible decay channels, obtaining  $\Gamma\sim 40 \,\GeV$ is  problematic: $8\,\TeV$ monojet searches \cite{Falkowski:2015swt}
 constrain the production cross section to be less than $\sim 0.5$pb (this bound is obtained by  recasting at the parton level the current experimental limits from \cite{Khachatryan:2014rra,Aad:2015zva}). As a result we are forced into the region with large values of $c_F(\hbox{TeV}/\Lambda)\gtrsim 80$ meaning very large multiplicities of the new charged particles at the TeV scale.

This problem is ameliorated if one instead considers the following process that could mimic a diphoton signal: $gg\ra s\ra a a\ra 4\gamma$, where the field $a$ is a very light scalar so that two photons from its decays are highly collimated and can be misidentified as a single photon.\footnote{The possibility that a pair of photons from a highly boosted particle could be identified as a single photon has previously been studied in \cite{Dobrescu:2000jt,Chang:2006bw,Toro:2012sv,Draper:2012xt,Ellis:2012zp,Ellis:2012sd,Curtin:2013fra}, and considered in the context of the $750 \,\GeV$ signal \cite{Knapen:2015dap,Agrawal:2015dbf,Chang:2015sdy,Bi:2015lcf,Chala:2015cev}.} In this construction the interaction between $s$ and $a$ can occur at tree, rather than loop, level so that effectively the model has large  values  of ``$c_F$''. It is natural for the field  $a$ to be very light if it is a pseudo Nambu Goldstone boson (PNGB) of a global symmetry, and the PNGB  structure does not forbid  
interactions of the form $a F \tilde F$.

In the rest of this paper we study this topology in detail, under the assumption that $a$ is the PNGB (i.e. the axion) associated with the breaking of a  global ${\rm U}(1)$ Peccei-Quinn (PQ) symmetry. Unlike the simple direct production model this allows a large width without problems from monojet or dijet limits. If the scale of PQ symmetry breaking is low, $\lesssim 400 \,\GeV$, the entire signal can be generated with a relatively modest number of additional fields charged under the Standard Model (SM) gauge group, but the theory is very close to strong coupling. At larger $f$ models can remain perturbative up to higher scales $\sim 10^7 \,\GeV$, with the width coming from invisible decays. One natural possibility we consider is that the invisible decays are to other PNGBs in a more complex hidden sector.

The structure of the paper is the following: we first present a model independent study of the mechanism in Section~\ref{sec:mi}; in Section~\ref{sec:em} we construct explicit models fitting the diphoton excess; in Section \ref{sec:wa} we compare the theories obtained to models of direct production; and in Section~\ref{sec:con} we discuss some additional possibilities and conclude. 

\section{Model independent analysis}\label{sec:mi}
A beneficial feature of the decay topology of interest is that many of the phenomenological features can be studied  in an effective field theory (EFT) approach below the scale of the global symmetry breaking. Assuming that the axion arises from the breaking of a ${\rm U}(1)$ PQ symmetry by a field $\phi$ we can expand it as 
\bea
\phi=\frac{f+s}{\sqrt{2}}e^{\frac{ia}{f}}~,~~|D_\mu \phi|^2\ra \frac{s (\partial_\mu a)^2}{f} ~.
\eea
Then the generic interactions important for the diphoton resonance are
\bea
{\cal L}=\frac{c_g\alpha_s}{12\pi}\frac{s}{f}G_{\mu\nu}G^{\mu\nu}+\frac{3\alpha}{4\pi}c_\gamma\frac{a}{f}\epsilon^{\alpha\beta\gamma\delta}F_{\alpha\beta}F_{\gamma\delta}+s\frac{(\partial_\mu a)^2}{f} ~.
\eea
In particular, all the relevant interactions can be parameterised in terms of three variables $c_g,c_\gamma,f$. To obtain phenomenologically viable models we also give the axion a small mass $m_a$, which is generated by a (technically natural) small explicit breaking of the PQ symmetry. 

\subsection{Production}
\begin{figure}
\begin{center}
\includegraphics[scale=0.5]{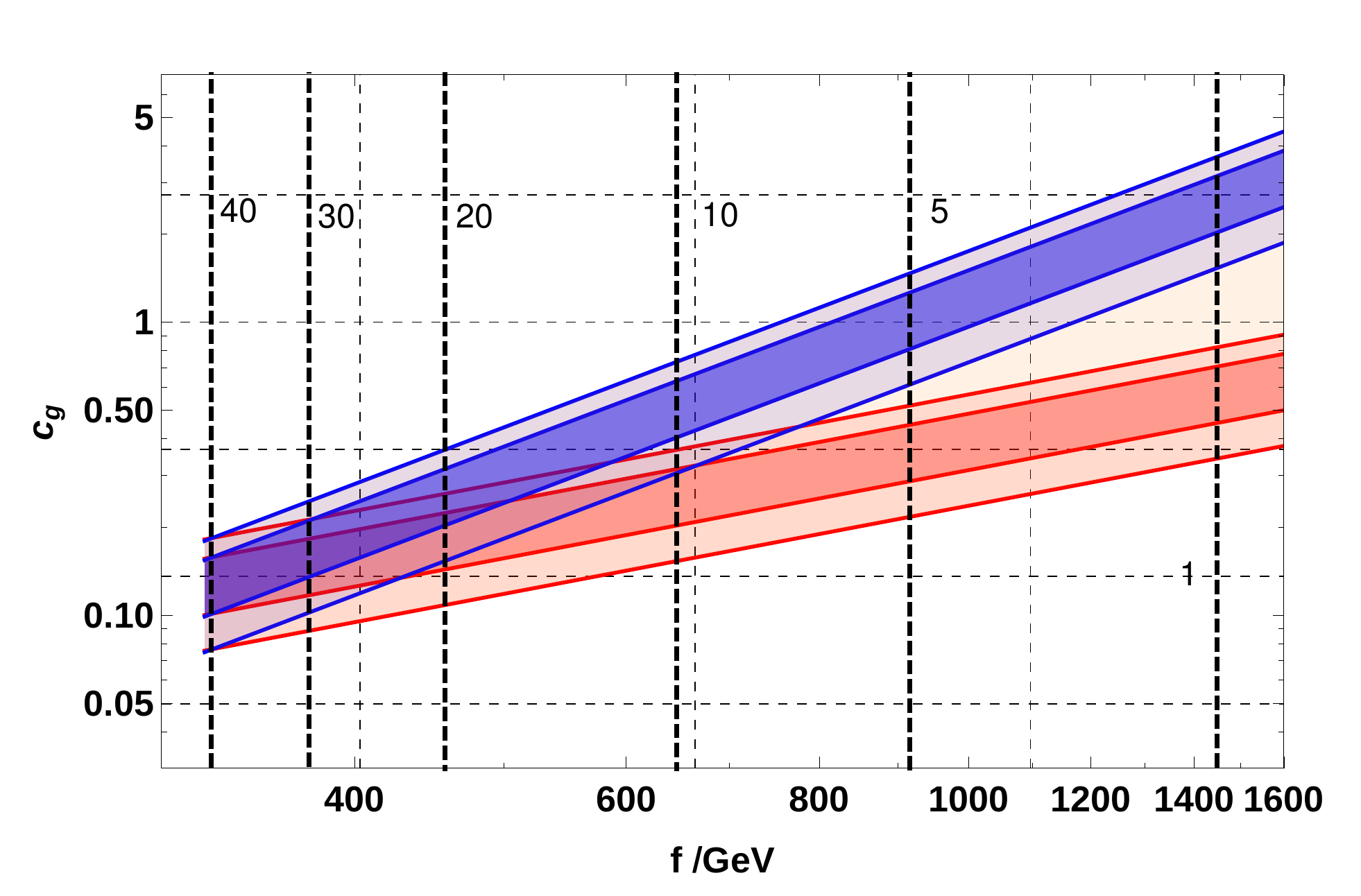}
\end{center}
\caption{\label{fig:ax_vs_f}$1,2 \sigma$ contours on the parameters of the axion model fit to the ATLAS excess for the cases where {\bf Red:} the invisible decay width is set to zero, and {\bf Blue:} the total decay width is set to $40$ GeV (due to invisible decay channels). Black vertical lines indicate the partial decay with into two axions in GeV. 
}
\end{figure}

The scalar $s$ is produced through gluon fusion and, to a good approximation, the cross section for this is
\bea \label{eq:prod}
&&\sigma(gg\ra s)\approx\frac{v^2}{f^2}\sigma(gg\ra h(m_h=750))\times\l|\frac{c_g }{F_{\triangle}(m_t,m_h=750)+F_{\triangle}
(m_b,m_h=750)}\r|^2 ,\nonumber\\
&&\sigma(gg\ra s) \simeq31 \,{\rm fb} ~ c_g^2  \l(\frac{\TeV}{f}\r)^2  ~,
\eea
where $F_{\triangle}$ are the usual fermion loop functions defined by Eq.~\eqref{eq:loopfunc} in Appendix~\ref{sec:appb}.
The partial decays width can be straightforwardly obtained 
 \bea
 \Gamma_{s\ra aa}=\frac{1}{32 \pi}\frac{M_s^3}{f^2}~,~~\Gamma_{a\ra\gamma\gamma}=\l(\frac{3 \alpha c_\gamma}{4\pi f}\r)^2\frac{m_a^3}{\pi} ~.
 \eea
Hence for  $s$ to  have a width to axions of $\sim 40 \,\GeV$ we need $f\simeq 320 \,\GeV$. Alternatively, higher values of $f$ can  give a significant width if  $s$ has additional decay channels and an invisible width $\Gamma_{inv}$. Assuming that the efficiency of photon identification is $\sim 100\%$ \cite{atlas}, and $\sim 85\%$ of axions successfully fake a photon, we obtain the number of events as a function of the PQ scale, both if the total width is raised with invisible decays, and also if there are no invisible decays (in which case the width is small unless $f$ is small) 
\bea
&&N_{ev}\approx \l\{\begin{array}{c} 
\displaystyle
7.5 ~ c_g^2  \l(\frac{\TeV}{f}\r)^4 \frac{40\,\GeV}{\Gamma_{inv}}~,~~\Gamma_{inv}\gg \Gamma_{s\ra aa}
 \\ 
 \displaystyle
 64~ c_g^2  \l(\frac{\TeV}{f}\r)^2 ~,~~ \Gamma_{s\ra aa}\gg\Gamma_{inv}~.
\end{array} \r. 
\label{eq:nevappr}
\eea
Since it is known that, at $2\sigma$, $N_{ev}\in [4,22]$ (for the excess reported by ATLAS)  this can be translated into a constraint on the parameters of the theory
\bea \label{eq:cggen}
c_g  \l(\frac{\TeV}{f}\r)^2 \in [0.73,1.78]~~\hbox{for}~~\Gamma_{inv}= 40 \,\GeV\nonumber~,\\
c_g  \l(\frac{\TeV}{f}\r) \in [0.23,0.55]~~\hbox{for}~~\Gamma_{inv}=0 \,\GeV ~.
\eea
In Fig.~\ref{fig:ax_vs_f} we plot the values of $c_g$  and $f$ compatible with the ATLAS observation.
From this we see that in the regime where $f$ is small (that is, the width comes from the decay of $s$ to $a$) we need quite a small coupling to gluons $c_g\sim 0.12$. In Section~\ref{sec:em} we show how this may arise from a model with additional heavy vector like quarks. In the large $f$ case the value of $c_g$ needed is fairly large, which can be naturally generated from models with new chiral quarks. Even when the width comes from invisible decays, monojet searches are much less constraining than for the direct production model. In the parameter regions of interest $\sigma_{p p \ra invisible}\lesssim 0.1$ pb, since typically $\Gamma_{s\ra aa} \gtrsim 1$ GeV.

\subsection{Axion faking a photon}
To reproduce the observed excess the two photons from the decay of each axion must be identified as a single photon. The ATLAS collaboration reports that the photons in the analysis are required to pass the ``tight photon'' tag which means meeting the selection criteria given in \cite{Aad:2010sp}. Although the photon event variables used by ATLAS are sophisticated, we can estimate that this roughly corresponds to requiring both photons hit the same plate of the first layer of ECAL, which has a very fine granularity in the $\eta$ direction
\bea \label{eq:etagran}
\Delta \eta \lesssim 0.0031 ~.
\eea
In Appendix~\ref{app:faking_photon} we show how this can be translated into a bound on the axion mass,
\bea
\label{eq:collphot}
m_a \lesssim 0.25 \,\GeV ~,
\eea
and also show that our simple analysis gives a good approximation to the result obtained from a more complete consideration of the actual cuts and detector properties.

At the same time we need the axion to decay before reaching the calorimeter (otherwise it would lead to displaced vertices, or escape the detector entirely). At ATLAS this is located $1.3\,{\rm m}$ from the interaction point. Therefore we obtain a constraint on $c_{\gamma}$, $m_a$, and $f$, which are related to the decay length of the axion in the lab frame by 
\bea
L_{{\rm Lab~frame}}\sim \frac{47}{c_{\gamma}^2} \left(\frac{f}{\TeV}\right)^2 \left(\frac{0.2\,\GeV}{m_a} \right)^4 \, {\rm m} \lesssim 0.6\,{\rm m} ~, 
\eea
where the decay length of $0.6~{\rm m}$ corresponds to an approximately $80\%$ probability that both axions decay before the ECAL. This gives
\bea
\label{eq:cgamma}
c_\gamma\gtrsim 9 \l(\frac{f}{\TeV}\r)\l(\frac{0.2 \,\GeV}{m_a}\r)^2 ~.
\eea  
\begin{figure}
\begin{center}
\includegraphics[scale=0.5]{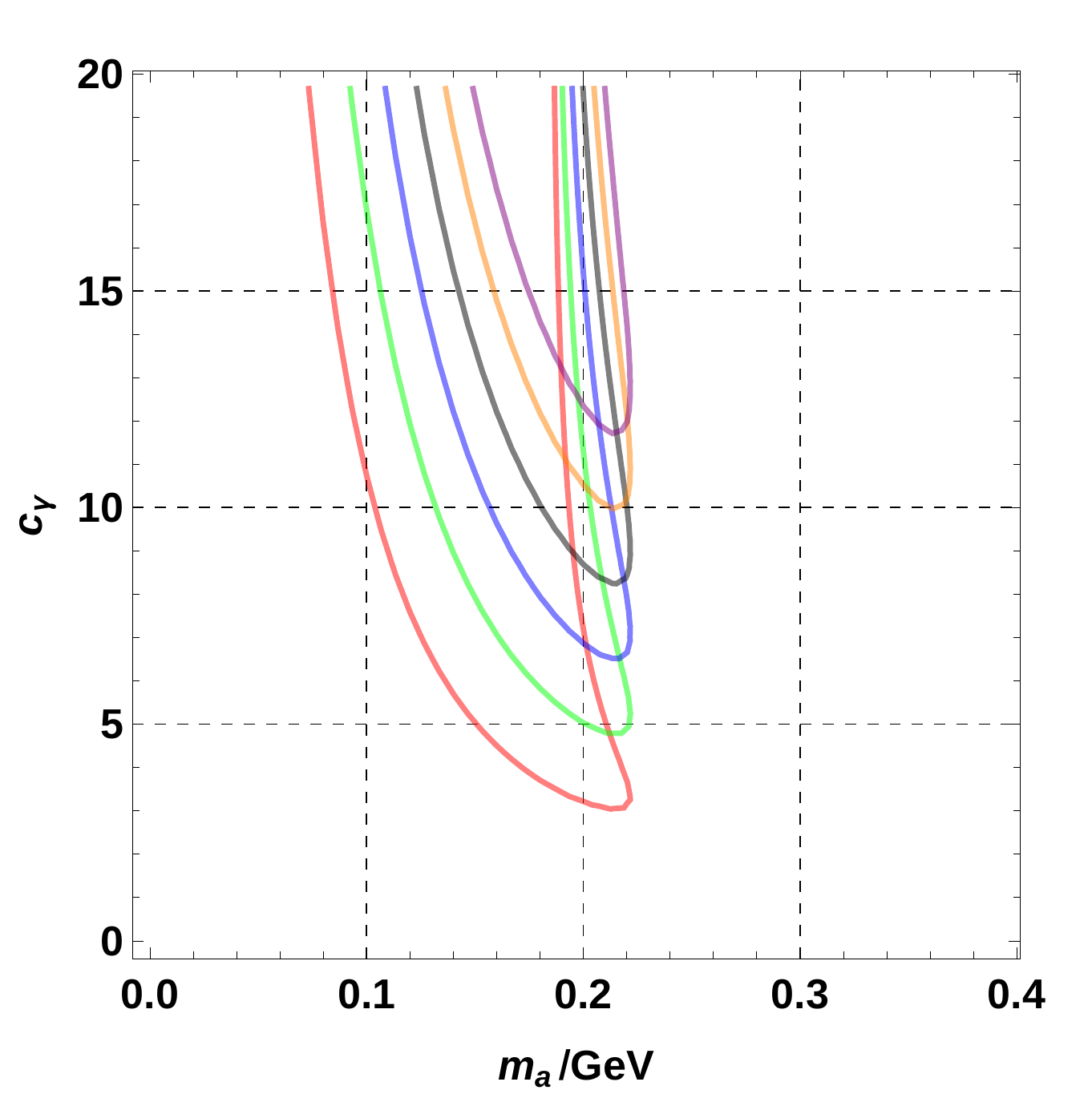}
\end{center}
\caption{\label{fig:ma_vs_cg}  Probability (P=0.6) contours that two axions will fake two photons as a function of the axion coupling to photons and the axion mass, for various values of $f$. Red, green, blue, black, orange, purple correspond to $f=350,550,750,950,1150,1350$ GeV.}
\end{figure} 
The viable parameter space, calculated following the procedure in Appendix \ref{app:faking_photon}, is presented in Fig.~\ref{fig:ma_vs_cg}, as a function of $c_\gamma$ and $m_a$  for different values of $f$. Demanding that 60$\%$ of axion pairs fake two individual photons constrains the axion mass to a relatively narrow window of $100 \div 200$ MeV, and $c_\gamma$ can be smallest when $f$ is small.

\subsection{Other constraints}
There are also other experimental constraints on light axion like particles. The couplings to photons in the parameter ranges of interest are close to being probed by beam dump experiments (other limits come from astrophysics and $e^+ + e^- \rightarrow \gamma + {\rm invisible}$ searches, but these are too weak to have any effect)  \cite{Jaeckel:2010ni}.
{
However, as argued in the previous section, the axion must have a decay length $l\lesssim 1~{\rm m}$ so cannot be constrained by the current NuCal\cite{Blumlein:1990ay} and future experiments  NA62 \cite{Krivda:2013ola}  and SHiP \cite{Anelli:2015pba,Alekhin:2015byh} \footnote{We would like to thank Oleksii Boiarskyi for pointing this out to us.}.}

QCD axions with decay constants $\sim \TeV$ are strongly excluded by the decay $K^+ \rightarrow \pi^+ a$ that looks like the rare SM process $K^+ \rightarrow \pi^+ \nu \nu$ \cite{Anisimovsky:2004hr}. In our models this does not automatically apply since, by construction, the axion typically decays inside the detector, resembling the much more common process $K^+ \rightarrow \pi^+ \pi^0$ albeit with a different pion mass. However, for simplicity and to be sure limits from this and similar processes are avoided, we mostly consider models where the extra SM charged states are such that $a$ has no anomalous coupling to gluons.

\section{Constructing explicit models}\label{sec:em}
Having discussed the generic features arising from the topology of interest we now consider models that can generate the required EFTs for $f$ in different ranges.

\subsection{Small $f$ models } \label{sec:smallf}
If $f$ is fairly small, $\lesssim 500 \,\GeV$, production and decay of $s$ requires that the parameters of the effective Lagrangian are roughly in the range 
\bea
&c_g\in[0.07,0.4]\nonumber\\
&c_\gamma\gtrsim 3 ~.
 \eea
 The large value of $c_\gamma$ relative to $c_g$ immediately points towards the states generating the axion coupling to photons being separate from those responsible for the coupling of $s$ to gluons. Further, stringent collider constraints on new coloured states mean that any coloured states that gain a mass dominantly from the vacuum expectation value (VEV) of $\phi$ would require a non-perturbatively large Yukawa coupling $\sim 1.5 \,\TeV / \left(250 \,\GeV\right) $.

Therefore we introduce two pairs of quarks that are vector like under the SM gauge groups and the PQ symmetry, but which couple to $\phi$ from cross terms. For example,
\bea
\label{eq:vec:quarks}
{\cal L}=M_1 Q_1^c Q_1 + M_2 Q_2^c Q_2 + y \phi Q_1^c Q_2 {+ \bar{y} \phi^{\dagger} Q_2^c Q_1} + {\rm h.c.} ~, 
\eea
with charges under the SM gauge group and ${\rm U}(1)_{PQ}$ of 
\begin{equation}
\begin{array}{cccccccc}
& {\rm SU(3)} & \times & {\rm SU}(2) & \times & {\rm U}(1)_Y & \quad {\rm U}(1)_\text{PQ}  \\
Q_1 \sim & 3 && 1 && Y & +1 \\
Q_1^c \sim & \bar 3 && 1 && -Y & -1 \\
Q_2 \sim & 3 && 1 && Y & 0 \\
Q_2^c \sim & \bar 3 && 1 && -Y & 0 
\end{array}
\end{equation}
where the electric charge is normalised as 
\bea
Q=T_3+Y ~.
\eea
These fermions generate a coupling of $s$ to gluons
\bea \label{eq:cgvl}
c_g=-\frac{y\bar y f^2}{M_1 M_2- y \bar y f^2/2} ~,
\eea
which, with $\mathcal{O}\left(1\right)$ Yukawa couplings and  $\mathcal{O}\left(\TeV\right)$ masses, can produce a $c_g$ of the correct size. By construction such fermions do not lead to a colour anomaly.

To create a sufficiently large coupling to photons we introduce extra chiral fermions uncharged under ${\rm SU}(3)$. For example, these could be chiral leptons that are singlets under ${\rm SU}(2)$ and have charge $Q_L$ under ${\rm U}(1)_Y$
\bea
{\cal L}=y_L\phi (L^c L)+ {\rm h.c.} ~,
\eea
where $L$ ($L^c$) have charge $-1$ ($0$) under the PQ symmetry, leading to a coupling
\beq
\label{eq:leptons}
c_\gamma=\frac{Q_L^2}{6} ~.
\eeq
In the case of $N$ copies of $L,L^c$ we need
\bea
\label{eq:nl}
\sqrt N Q_L \gtrsim 7.3\sqrt{\frac{f}{\TeV}}\frac{0.2 \MeV}{m_a} ~.
\eea               


To obtain models without a very large number of new states we need extra leptons with charges of at least $+2$.  As they can be produced by the electroweak Drell-Yan processes the LHC has set fairly stringent limits on the masses of such states. In particular, if the new leptons are stable on collider scales and have charge $+2$ the current bound is roughly $\gtrsim 700$ GeV \cite{Chatrchyan:2013oca}. Since we require a multiplicity of leptons $N \sim 4$ this translates into $M\gtrsim 850 \,\GeV$ (by considering PDFs and keeping the number of signal events fixed). In the case of $Q_L=4,N=1$ and we need $M_l\gtrsim 800 \,\GeV$.

Alternatively  the leptons could decay promptly inside the collider. For charge $+2$ leptons this can occur through a dimension six operator, for example
\bea
(\bar L\gamma_\mu e_R)((H^c)^\dagger \dblarrow D_\mu H) ~,
\eea
which couples $W^+$ to the right handed current $(\bar L\gamma_\mu e_R)$. This leads to a same sign dilepton final state, which is constrained by searches for doubly charged Higgs bosons \cite{ATLAS:2014kca}. Taking into account the  leptonic branching fraction of $W$ we see that doubly charged leptons with mass $m_L\gtrsim 600-700 \,\GeV$ will be consistent with the current limits. Allowing only  couplings to $\tau$s will further reduce the bound. We can estimate it  by noting that the $\tau$ decays leptonically $\sim 40\%$ of the time, so that the constraint on the lepton mass becomes $\sim 600$ GeV. Searches for charge $5/3$ top partners also lead to similar limits when translated to our models.

\begin{figure}
\begin{center}
\includegraphics[scale=0.75]{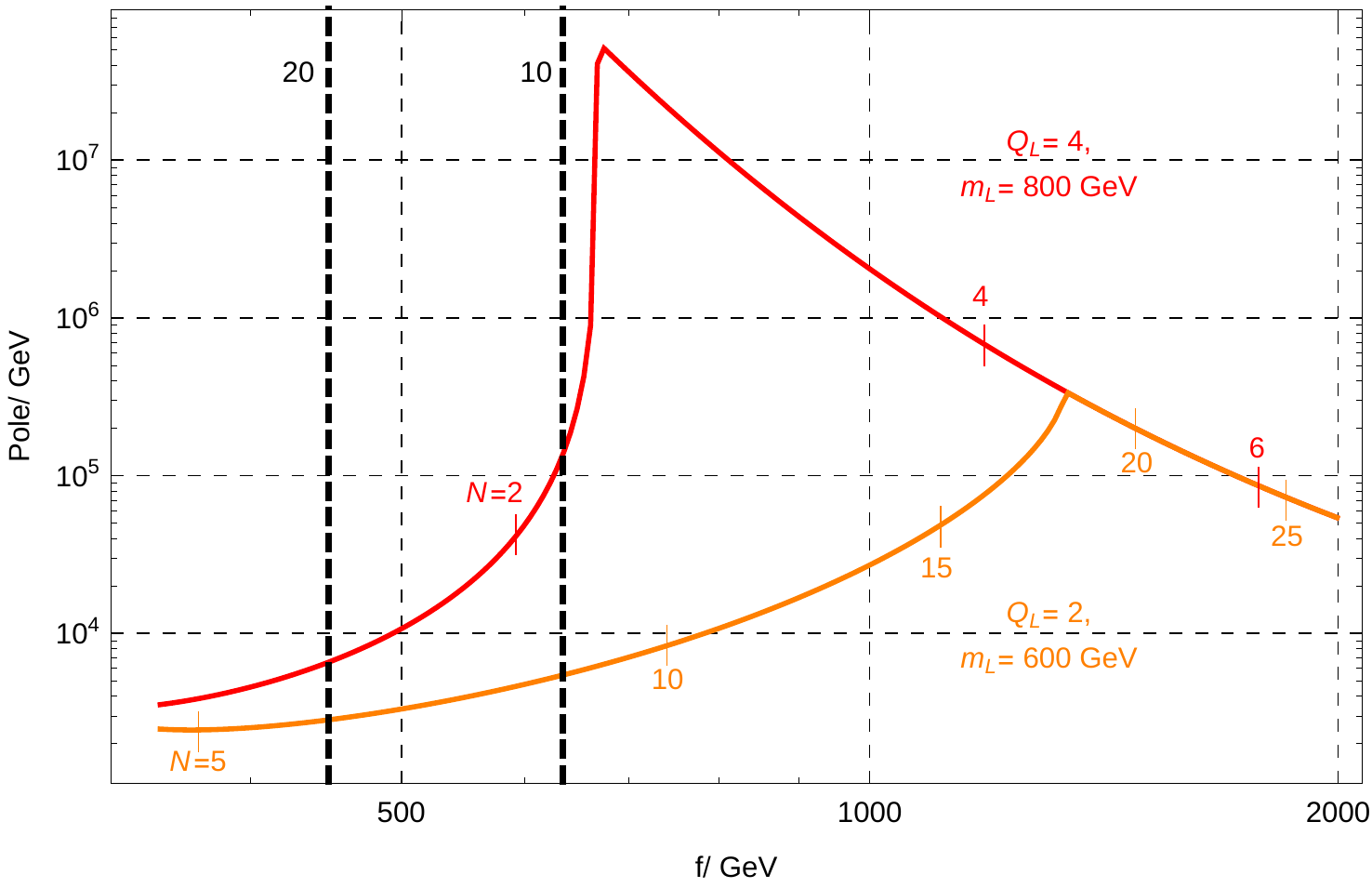}
\end{center}
\caption{\label{fig:pole} The scale of strong coupling as a function of $f$, with the number of new leptons $N$ varied to satisfy Eq.~\eqref{eq:cgamma}, and labeled on the curves. We show results for the two sets of lepton masses and charges compatible with direct LHC searches.  For small $f$, strong coupling occurs due to the scalar quartic coupling, so the scale of this increases with increasing $f$. For large $f$ strong coupling is due to the ${\rm U}(1)$ hypercharge gauge coupling, so occurs at lower scales as $f$ is increased and the $c_\gamma$ required grows. We also show the values of $f$ corresponding to a width to axions $\Gamma_{s\rightarrow a a} = 20,10\,\GeV$ by black vertical lines.}
\end{figure}

Models with small $f$ are typically close to being strongly coupled. First, the new leptons will make the ${\rm U}(1)_Y$ gauge coupling $g_Y$ run faster than in the SM. Also, these leptons need a large Yukawa coupling to get a sufficiently large mass. Finally, the small value of $f$ relative to $750\,\GeV$ means that the quartic coupling of $s$ is big. For the model with N pairs of singlet leptons with charge $Q_L$, the one loop renormalisation group equations for these couplings are
\bea \label{eq:running}
\frac{d y_L}{dt}=\frac{1}{16\pi^2}\l[ y_L^3(N+1)-6 g_Y^2 y_L Q_L^2 \r]\nonumber\\
\frac{d\lambda}{d t}=\frac{1}{16\pi^2}\l[10\lambda^2+4 N y_L^2 \lambda - 4 N y_L^4 \r]\nonumber\\
\frac{d g_Y}{dt}=\frac{g_Y^3}{16\pi^2}\l[\frac{41}{6}+\frac{4}{3} N Q_L^2\r] ~,
\eea
where $M_s^2=\lambda f^2$, and $m_l=y_L f/\sqrt{2}$. In Fig.~\ref{fig:pole} we show the scale of strong coupling (defined as any coupling $= 4\pi$, or the quartic running negative) as a function of $f$ from a numerical solution of these equations. The charge of the leptons is fixed to be equal to $2$ or $4$ and the number of species $N$ is taken to be minimal satisfying the inequality Eq.~\eqref{eq:cgamma} for a sufficiently prompt axion decay
\bea
N\gtrsim \frac{0.054}{Q_L^2}\frac{f}{\GeV} ~.
\eea

We see that models with small PQ breaking generically have a very low cut off scale, 
and, in the case when the partial decay width into two axion is $40\,\GeV$, our model is already 
strongly coupled at the TeV scale.  However, as mentioned in Section~\ref{sec:wa}, in this case we can easily explain the observed width with $N= 4,Q_L=2$.

For $f$ slightly larger, $\sim 500\,\GeV$, perturbativity can be maintained up to $10^4 \,\GeV$, while keeping a width $\Gamma_{s\rightarrow a a}$ of $15\,\GeV$. Given the current uncertainties on the signal this may be regarded as sufficient. Alternatively, the width can be increased to $40\,\GeV$ by including hidden sector fermions, uncharged under the SM, with PQ preserving couplings $y_{\psi} \phi \bar{\psi}_h \psi_h$ that give the fermions masses. These give an invisible width 
\beq
\Gamma_{inv}=\frac{N}{8\pi} M_s\l(1-\frac{2 y_{\psi}^2 f^2}{M_s^2}\r)^{3/2} y_{\psi}^2 ~,
\eeq 
where $N$ is the number of extra vector-like fermions. With 4 pairs of vector-like hidden sector fermions a total width of $40\,\GeV$ can be obtained for $f=510 \,\GeV$. The invisible decays mean the production cross-section must be correspondingly increased, however considering Eq.~\eqref{eq:cggen} and Eq.~\eqref{eq:cgvl} it can be seen that this is not problematic.

\subsection{Large $f$ models } \label{sec:lf}
In the opposite regime when $f$ is large, taking for definiteness, $f \sim 1.2 \,\TeV$ and $\Gamma_{inv}=40$, we require
\bea
c_g\in[1.1,2.6]~,~~c_\gamma\gtrsim 11 ~,
\eea
so much larger charges/multiplicities are needed to have the correct value of $c_\gamma$. Since larger values of $c_g$ are required for production it is possible for a single set of chiral PQ charged fermions to give both $c_\gamma$ and $c_g$ simultaneously. The larger VEV also allows the coloured chiral fermions to be safely outside LHC limits without non-perturbatively large Yukawa couplings. Consider, for example, the model
\bea
{\cal L}=\phi Q_1^c Q_1 +\phi^\dagger Q_2^c Q_2 + {\rm h.c.}~,
\eea
where the quantum numbers of the fields under the SM gauge group and the ${\rm U}(1)_{PQ}$ are
\begin{equation}
\label{eq:largef}
\begin{array}{cccccccc}
& {\rm SU(3)} & \times & {\rm SU}(2) & \times & {\rm U}(1)_Y & \quad {\rm U}(1)_\text{PQ}  \\
Q_1 \sim & 3 && 1 && Y_1 & -1 \\
Q_1^c \sim & \bar 3 && 1 && -Y_1 & 0 \\
Q_2 \sim & 3 && 1 && Y_2 & +1 \\
Q_2^c \sim & \bar 3 && 1 && -Y_2 & 0 
\end{array}
\end{equation}
These generate couplings
\bea
c_g=2~,~~c_\gamma=\frac{Y_1^2-Y_2^2}{2} ~.
\eea
Consequently we need fields with the electric charge larger than 4, or we can again introduce chiral leptons to increase $c_\gamma$.

The decay width to axions is too narrow to account for the experimental observations, therefore we need a large invisible width, which requires additional light states coupled strongly to $s$. Simply introducing extra scalars or fermions to the theory does not automatically help, since fields with large couplings to $\phi$ are expected to gain a mass $\sim f$ and decays of $s$ to these are kinematically forbidden. 

One option is to tune a state light (in the next section we discuss an alternative of a model with multiple PNGBs). For example adding a new scalar $\eta$ with a bare mass and a coupling to $\phi$ of $\lambda \left|\phi\right|^2 \eta^2$, gives an invisible width 
\bea
\Gamma(s\ra \eta\eta)=\frac{1}{8\pi}\frac{\lambda^2 f^2}{M_s}\sqrt{1-\frac{4 m_\eta^2}{M_s^2}} ~.
\eea
Similarly a fermion can be made light by introducing two fermions $N_{1,2}$ with opposite PQ charges. Then the vector-like mass term $M_{12}  \l(i N_2^T\sigma^2 N_1+ {\rm h.c.}\r)$ can be tuned against the mass induced by a coupling to $\phi$ of $y_1 \phi \l(i N_1^T\sigma^2 N_1+ {\rm h.c.}\r)+ \left(N_1 \leftrightarrow N_2\right)$. This gives an invisible width of
\bea
\Gamma(s\ra n_1 n_1)=\frac{1}{8\pi} M_s\l(1-\frac{4 m_{n1}^2}{M_s^2}\r)^{3/2} y_1^2 ~.
\eea

Despite needing more new SM charged matter than if $f$ is small, the large $f$ case can remain perturbative to a higher scale. The quartic coupling of $s$ is small since $M_s < f$, and the Yukawa couplings of the new quarks do not need to be especially large. Instead strong coupling first occurs due to the ${\rm U}(1)$ hypercharge gauge coupling. Since the change in the beta function is proportional to the sum of the hypercharges of the states $Y_1^2+Y_2^2$, while $c_{\gamma}$ is proportional to $Y_1^2-Y_2^2$, the best chance to obtain a weakly coupled theory is to have $Y_2=0$. Then the change in the ${\rm U}(1)_Y$ beta function can be related to the anomaly coefficient by $\Delta \beta_Y = 8 c_{\gamma}$. For $f\sim \TeV$, models can remain weakly coupled until $\sim 10^6\,\GeV$  as shown in Fig.~\ref{fig:pole}.

\subsection{Invisible width from a larger coset} \label{sec:naxion}

A large invisible width can occur without tuning in models with many PNGBs. Suppose that the hidden sector contains $n$ pairs of quarks, coupled to $n^2$ complex scalar fields, $\phi_{ij}$, by a term
\beq \label{eq:naxq}
\lambda \phi_{ij} Q^c_i Q_{j} +\tilde \lambda \phi_{ij}^* \tilde Q^c_i \tilde Q_{j} +{\rm h.c.} ~.
\eeq
Then the theory has a ${\rm U}(n)\times {\rm U}(n)={\rm SU}(n)\times {\rm SU}(n) \times {\rm U}(1)_A \times {\rm U}(1)_V$ global symmetry, with the axial ${\rm U}(1)_A$ anomalous. We assume the potential of the theory is such that (after rotating in field space) the diagonal scalars $\phi_{ii}$ all get approximately equal VEVs. This spontaneously breaks the global symmetry to the ${\rm SU}(n)_V\times {\rm U}(1)_V$ subgroup, and there are $n^2$ PNGBs. The scalar sector of the theory can be expanded as
\beq
\hat \phi = \frac{s+\hat{\rho}+F}{\sqrt{2n}} \dblone \times e^{i  \hat{\Pi}/F}\times e^{i a/F} + {\rm h.c.} ~.
\eeq
Here $\hat \Pi$ and $a$ are the PNGB associated with ${\rm SU}(n)$ and ${\rm U}(1)$ breaking, $s$ is the corresponding Higgs field along the diagonal direction, which will be the $750\,\GeV$ state, and $\hat{\rho}$ are the remaining $n^2-1$  massive scalars.
Out of the PNGBs only $a$ will have an anomalous coupling to photons, since
 \bea
 {\rm Tr}\left( T^a Y^2 \right)=0~,
 \eea 
 where $T^a$ are the generators of ${\rm SU}(n)$.  Similarly the couplings of the scalars $\hat \rho$ to gluons and photons will also vanish. The $n^2$ PNGBs couple equally to $s$, so its total width, and branching ratio to axions, is
 \beq 
 \label{eq:width}
\Gamma_s = \frac{n^2}{32 \pi} \frac{M_s^3}{F^2}~,~~{\rm Br}(s\ra aa)=\frac{1}{n^2} ~.
\eeq
The couplings of $s$ and $a$ to gluons and photons respectively are given by (analogously to  Eq.~\eqref{eq:largef})
\bea
\frac{c_g}{f} =\frac{2n}{F} ~, ~~\frac{c_\gamma}{f}=\frac{n}{F}\l(\frac{Y_1^2-Y_2^2}{2}\r) ~.
\eea
Therefore, using Eq.~\eqref{eq:nevappr}, we find that the observed number of events can be explained if $F \sim 4\,\TeV$. Fitting the width we find $n\sim 12$, and the hypercharge must be $Y_1\gtrsim 2.5$ if $Y_2=0$. Since the other scalars couple only weakly to the Standard Model, they do not violate any existing observations.

Similarly we can extend the small $f$ models (see Section~\ref{sec:smallf}) to the ${\rm U}(n)\times {\rm U}(n)$ construction with a Lagrangian
\bea
\label{eq:vec:quarksN}
&{\cal L}=M_1 Q_1^{c, i} Q_1^i+ M_2 Q_2^{c, i} Q_2^i+ y \phi_{ij}  Q_1^{c,i} Q_2^j {+ \bar{y} \phi^{*}_{ij} Q_2^{c, j} Q_1^i} \nonumber \\ 
&+y_L\phi_{ij} (\tilde L_i^c L_j)+ {\rm h.c.}\nonumber ~.
\eea
The couplings to  gluons and photons are given by
\bea \label{eq:cgvl}
\frac{c_g}{f} &&=-\frac{y\bar y F }{M_1 M_2-\frac{y \bar y F^2}{2n}}~,\\
\frac{c_\gamma}{f} &&=\frac{n Q_L^2}{6 F} ~.
\eea
Requiring the width is $\Gamma=40\,\GeV$, we obtain a constraint $F/n=320 \,\GeV$. Further, for prompt axion decay $Q_L\gtrsim 4$ is needed, independently of $F$ and $n$, and the appropriate values of $c_g$ can be obtained for $y\, \bar y \sim 1$. Reasonable points in parameter space are possible, for example $n=3$ and $F,M_{1,2}\sim \TeV$.

Although in our conventions $F$ can be large, the running of the quartic scalar coupling is enhanced relative to  Eq.~\eqref{eq:running} by the number of PNGBs. The exact form of the beta function depends on the scalar potential. For example, if the quartic dominantly comes from a term $V\sim \lambda {\rm Tr}\left(\phi^{\dagger}\phi\phi^{\dagger}\phi\right)$, it has a dependence on $n$ of $\beta_{\lambda} \sim  \left(4n+6 \right)\lambda^2$. To compare with the $n=1$ small $f$ case, we redefine the quartic $\lambda' = \left(4n+ 6\right) \lambda$, so that the running of $\lambda'$ is independent of $n$ at one loop. The mass of $s$ is related to $\lambda$ by $M_s^2 = F^2 \lambda /n$, therefore the value of $\lambda'$ is related to the width by
\beq
\lambda' = 32\pi \frac{\Gamma_s}{M_s} \left(4+ \frac{6}{n} \right)~.
\eeq
We see that increasing $n$ decreases the required value of $\lambda'$ by an order one amount. As a result strong coupling can occur at a higher scale, and for $n\sim 3$ the theory could remain weakly coupled up to scales $\sim 10^6~\GeV$.

\section{Comparison with the direct diphoton production}
\label{sec:wa}

It is interesting to compare the requirements on UV completions in direct diphoton production and diaxion models. In particular, this allows us to determine if anything has been gained from the extra structure introduced. 

Let us start by looking at the axion construction in the small $f$ regime. 
In the  $f \lesssim 400 \,\GeV$ limit the setup provides a nice explanation of the width of the diphoton signal without introduction of new invisible decays. In contrast this seems to be almost impossible in the models parametrized by the Lagrangian  Eq.~\eqref{eq:noax}. Unfortunately  these  nice features are accompanied by still moderately large multiplicities of new particles  $\sqrt{N} Q_L\gtrsim 4.6$, and strong coupling at the TeV scale. 

Both the diaxion in the large $f$ regime and the diphoton construction, require invisible 
decay channels, as well as large multiplicities of SM charged particles, although for 
different reasons. In the direct diphoton construction they are needed to fit the 
observed signal as well as avoid bounds from invisible decay searches, while for diaxions they are needed for prompt axion decays. We can compare the needed multiplicities of new fields for the two setups by assuming the interactions with photons are both generated by $N$ species of charge $Q$ chiral fermions. Then the numerical values of the effective couplings are related by
\bea
& \displaystyle c_\gamma=\frac{N Q^2}{6}~,~~ c_F=\frac{3NQ^2}{4}~,\\
& \displaystyle \implies c_F=\frac{9}{2}c_\gamma~,
\eea
where we have set the cutoff scale of Eq.~\eqref{eq:noax} to be $\Lambda\equiv f$. From Eq.~\eqref{eq:cgamma}, for the axion model to work we need a theory that generates couplings
\bea
\label{eq:cfest}
c_F\l(\frac{\hbox{TeV}}{f}\r)=\frac{9}{2}\, c_\gamma \l(\frac{\hbox{TeV}}{f}\r)\gtrsim40 \left(\frac{0.2\,\GeV}{m_a}\right)^2 ~.
\eea
In contrast, for the direct production model to match the signal while satisfying the monojet constraints $c_F\l(\TeV/f\r)\gtrsim 80$ is needed. Comparing this estimate with Eq.~\eqref{eq:cfest} we  see that the axion allows the multiplicity of new particles to be reduced by roughly a factor of $\sim2 \left(m_a/0.2\,\GeV\right)^2$. 
The improvement seems to be almost marginal, however note that this value depends strongly on the collimated photon condition $m_a\lesssim 0.25$ (see Eq.~\eqref{eq:collphot}).
If our estimate turns out to be over-constraining  and e.g.  $m_a\lesssim 0.4$ GeV can be tolerated, a reduction of 
 $\sim10$  in the multiplicity of new particles is possible. 
 {
 If a large width is not confirmed models with $f\gtrsim 700 $ GeV (see Fig.\ref{fig:ax_vs_f}) can still provide a viable description of the signal, since the experimental resolution is around $6$ GeV. However, unlike models with a direct decay to diphotons where the signal can be fitted with moderate couplings to photons (see Fig.\ref{fig:noax}), the prompt axion decay condition still requires a large coupling to photons (see Fig.\ref{fig:ma_vs_cg}), 
\bea
&&\hbox{diphoton: } c_F\l(\frac{\hbox{TeV}}{f}\r)\sim 4,\nonumber\\
&&\hbox{diaxion: }c_F\l(\frac{\hbox{TeV}}{f}\r)=\frac{9}{2}\, c_\gamma \l(\frac{\hbox{TeV}}{f}\r)\gtrsim 30 ~,
\eea
reducing the appeal of the diaxion construction.
}


\section{Conclusion} \label{sec:con}

In conclusion let us summarise the main results of our paper. We have considered the possibility that the recent ATLAS results are due to the decay of a 750 GeV resonance into two axions which subsequently decay to two pairs of highly collimated photons. The conditions and constraints on such a scenario have been discussed in a model independent EFT framework, and we have shown that the tentative 
 signal width can be explained. However, for the axions to be sufficiently collimated while also decaying before reaching the calorimeter, we need the axion mass to be in a narrow window $150 \div 200$ MeV.

A drawback of our analysis, which is crucial for identifying the viable parameter 
space, is the lack of a full detector simulation of the conditions for axion decays to fake individual photons. Although we have made conservative estimates, and reproduce the 
results of more sophisticated simulations well \cite{Draper:2012xt}, our results should therefore be taken with 
a pinch of salt. 

\begin{figure}
\begin{center}
\includegraphics[scale=0.7]{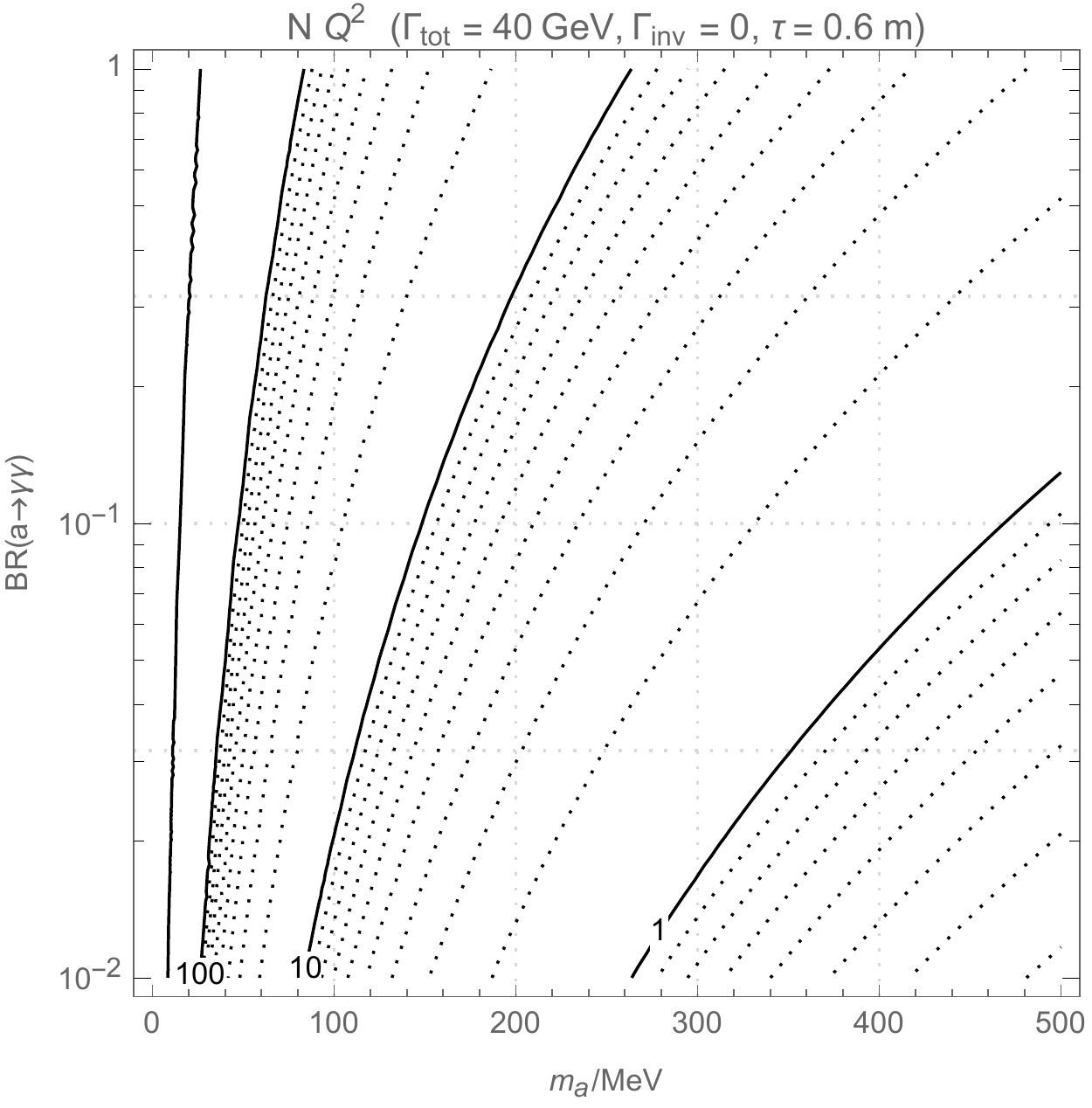}
\caption{\label{fig:mabr}The isocontours of the minimal number of the fields and charges $N Q^2$ needed to ensure a prompt axion decay $\tau \lesssim 0.6 $ m as a function of the axion mass and its branching into two photons for $f=320 $ GeV.}
\end{center}
\end{figure}

There are also some model building questions that we have not considered but need to be addressed in a complete theory. For example, the new chiral matter needs a mechanism to 
decay. This is not a major obstacle, and can occur via a PQ preserving coupling to SM matter. 
An explicit source of PQ breaking must also be introduced so that the axion gets an 
appropriate mass. It is also interesting to consider UV completions, especially since strong 
coupling is often not far from the TeV scale.  It may also be possible to connect the model to the SM flavour structure, suppressing the Yukawa couplings of the light SM fields with the PQ 
symmetry. Although we have not considered it in the present work, models where the axions 
have some invisible decay channels, e.g. into light hidden sector gauge bosons, might be possible, and could relax 
the prompt decay requirement (see Fig. \ref{fig:mabr}).

A phenomenological feature of our model is that unlike many other 
explanations of the diphoton anomaly it does not predict any excess in the ZZ and WW final states. Therefore if such a signal is observed the topologies we consider will be strongly disfavoured. On the other hand, if no excesses are seen in these channels (and the diphoton anomaly persists) the motivation for the theories we have considered will improve significantly. Similarly if the diphoton signal remains and the width is indeed large, this will also provide reason to further consider these type of models. Also, the presence of two photons instead of one increases the chance of production of an $e^+ e^-$ pair in the detector tracking system, which would be identified as a converted photon. It is an interesting question for future work whether this generic prediction allows collimated photons to be efficiently distinguished from real diphoton signals.

\section*{Acknowledgements}
We are grateful to Giovanni Villadoro for very useful discussions. The work of A.R.\ is supported by the ERC Advanced Grant no. 267985�``DaMESyFla''.

\appendix

\section{Loop functions} \label{sec:appb}
For completeness the loop functions for the production of $s$ used in Eq.~\eqref{eq:prod} are \cite{Ellis:1975ap,Shifman:1979eb}
\bea
\label{eq:loopfunc}
F_{\triangle}(m_f,M_s)=\frac{3}{2\tau^2}\l[ \tau+(\tau-1)f(\tau)\r]\nonumber\\
\tau=\frac{M_s^2}{4 m_f^2},~~f(\tau)=\l\{\begin{array}{c}
\hbox{arcsin}^2\sqrt{\tau},~~\tau\leq1\\
-\frac{1}{4}\l[\log \frac{1+\sqrt{1-1/\tau}}{1-\sqrt{1-1/\tau}}-i \pi\r]^2 ~.
\end{array}\right.
\eea

\section{Mistaking axions for photons}
\label{app:faking_photon}
Considering the decay of an axion with velocity $v$ into two photons, the total angle between the two photons in the lab frame $\theta_B$ is related to that in the axion rest frame $\theta$ by
\bea
\cos\theta=\pm\sqrt{\frac{2v^2-1-\cos\theta_B}{v^2(1-\cos\theta_B)}} ~.
\eea
Then the probability that the angle between two photons will be less than $\theta<\theta_M$ will be given by the integral
\bea
\int_{\theta_X}^{\pi/2}\sin\theta \,d\theta=\cos\theta_X ~,
\eea
where
\bea
\cos\theta_X=\sqrt{\frac{2v^2-1-\cos\theta_M}{v^2(1-\cos\theta_M)}} ~.
\eea
The distance between the collision point and the  calorimeter is of the order of $L_D=O(1\,{\rm m})$ and we need the axion to  decay before reaching the calorimeter plates. The plates of the liquid argon calorimeter have a minimum angular separation $\delta\eta_{ATLAS}=0.0031$ and much larger granularity in the $\Delta\phi$ direction. The reference \cite{Draper:2012xt} studied a probability of faking a photon by an axion for the Higggs decay. They found that the complete ATLAS analysis used for the tight photon selection, which utilizes more sophisticated variables, effectively means that the  cut on the $\delta \eta$ separation between the photons to be 
\bea
\delta \eta_{\gamma\gamma}< \frac{1}{2}\delta \eta_{ATLAS}~,~~\delta \eta_{ATLAS}=0.0031 ~.
\eea
Instead of a putting a cut only on the $\delta \eta$ direction between the photons we impose the conservative requirement that total angular separation between the photons be less then 
\bea
\delta R_{\gamma\gamma}<  \frac{1}{2}\delta \eta_{ATLAS} ~. 
\eea
Then, for an axion proper decay time $\tau$ and boost factor $\gamma=1/\sqrt{1-v^2}$, the total probability of two photons having a spacial separation less than $\delta d < \delta \eta_{MIN} L_D$ is
\bea
\int_0^{L_D}\frac{1}{\gamma\tau}e^{-\frac{x}{\gamma\tau}}\sqrt{\frac{2v^2-1-
\cos\l(\frac{\Delta\eta_{MIN} L_D}{L_D-x}\r)}{v^2(1-\cos\l(\frac{\Delta\eta_{MIN} L_D}{L_D-x}\r))}}\, dx ~.
\eea
To validate our approximations we have compared the results obtained using our technique for the SM Higgs decay to the results of \cite{Draper:2012xt} who performed a more complete simulation accounting for the varying granularity over the detector. We find our results match theirs fairly accurately, and we are slightly more conservative on the maximum mass possible to resemble a photon.

\begin{figure}
\begin{center}
\includegraphics[scale=0.7]{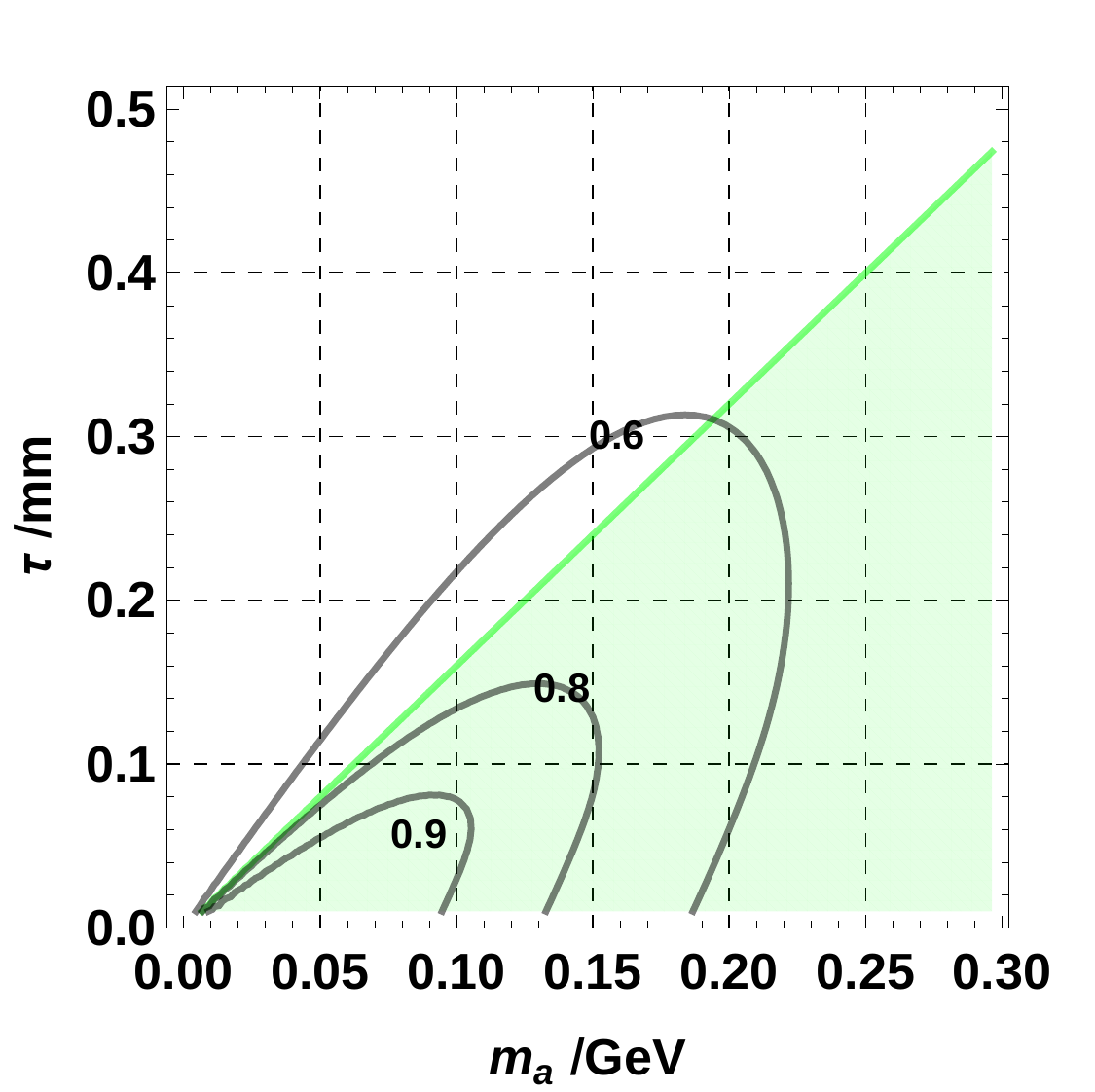}
\end{center}
\caption{\label{fig:prob} Probability contours to detect two axions as two photons, the green region indicates the parameter space where the decay length is less than  $0.6$ m.}
\end{figure}

Now  we proceed to the decay of the 750 resonance. Since the boost of the state $s$ is negligible, the axion boost is $375\GeV/m_a$. The results are presented in Fig.~\ref{fig:prob} where we show the probability to mistag the four photons from two axions as two photons as a function of the axion mass. In order for a reasonable proportion to be misidentified the decay length in the lab frame should be roughly less than 1 meter and the axion should be in the region of $\sim O(100)\,\MeV$.

\end{document}